\documentclass[Article,twocolumn,showpacs,showkeys,amsmath,amssymb,superscriptaddress,prl]{revtex4-1}
\usepackage{graphicx}
\usepackage{dcolumn}
\usepackage[dvips]{color}
\usepackage{bm}
\usepackage{lettrine}
\usepackage{subfigure}
\usepackage{type1cm}
\begin{document}
\title{Pressure induced multicriticality and electronic instability in quasi-kagome ferromagnet URhSn}
\author{Arvind Maurya}
\email{arvindmry@imr.tohoku.ac.jp}
\affiliation{Institute for Materials Research, Tohoku University, Oarai, Ibaraki 311-1313, Japan }
\author{Dilip Bhoi}
\affiliation{Institute for Solid State Physics, The University of Tokyo, Kashiwa, Chiba, 277-8581, Japan}

\author{Fuminori Honda}
\affiliation{Institute for Materials Research, Tohoku University, Oarai, Ibaraki 311-1313, Japan }
\author{Yusei Shimizu}
\affiliation{Institute for Materials Research, Tohoku University, Oarai, Ibaraki 311-1313, Japan }
\author{Ai Nakamura}
\affiliation{Institute for Materials Research, Tohoku University, Oarai, Ibaraki 311-1313, Japan }
\author{Yoshiki J. Sato}
\affiliation{Institute for Materials Research, Tohoku University, Oarai, Ibaraki 311-1313, Japan }
\author{Dexin Li}
\affiliation{Institute for Materials Research, Tohoku University, Oarai, Ibaraki 311-1313, Japan }
\author{Yoshiya Homma}
\affiliation{Institute for Materials Research, Tohoku University, Oarai, Ibaraki 311-1313, Japan }
\author{M. Sathiskumar}
\affiliation{Institute for Solid State Physics, The University of Tokyo, Kashiwa, Chiba, 277-8581, Japan}
\author{Jun Gouchi}
\affiliation{Institute for Solid State Physics, The University of Tokyo, Kashiwa, Chiba, 277-8581, Japan}
\author{Yoshiya Uwatoko}
\affiliation{Institute for Solid State Physics, The University of Tokyo, Kashiwa, Chiba, 277-8581, Japan}
\author{Dai Aoki}
\affiliation{Institute for Materials Research, Tohoku University, Oarai, Ibaraki 311-1313, Japan }

\date{\today}

\begin{abstract}

\textbf{We report an unconventional class of pressure induced quantum phase transition, possessing two bicritical points at 6.25 GPa in URhSn. This unique transformation accompanies a Fermi surface reconstruction, demarcating competing ordered phases suitably described with localized and itinerant description of the magnetic $5f$-electrons. Ferromagnetic fluctuations over a wide range of temperatures and pressures in the pressure-induced low temperature phase  are evidenced by a robust $T^{5/3}$ temperature dependence of resistivity up to 11 GPa, which is a characteristic of elusive marginal Fermi-liquid state.}

\end{abstract}

\maketitle

Pressure is proven to be an effective knob for tuning lattice coupled competing interactions and hence reveal exotic phases and associated interesting phenomena which otherwise would have been inaccessible. Suppression of an antiferromagnetic order by pressure often leads to emergent unconventional superconductivity or non-Fermi liquid behaviour in vicinity of a putative quantum critical point (QCP), maneuvered by quantum fluctuations ~\cite{Hilbert2007RMP,Pfleiderer2009RMP,Stewart 2001 RMP NFL}). However, in the case of ferromagnetic ground state a generic second order QCP is usually avoided by coupling of the order parameter to fermionic soft modes~\cite{Belitz}. This results in either appearance of a modulated magnetic structure or switching to a first order transition above a pressure specified by a tricritical point~\cite{Belitz, Brando}. So far, only handful of materials, namely YbNi$_4$(P$_{1-x}$As$_x$)$_2$~\cite{Steppke 2013Science YbNi4P2}  and CeRh$_6$Ge$_4$~\cite{Shen2020Nature CeRh6Ge4, kotegawa 2019JPSJ CeRh6Ge4} manifest ferromagnetic QCP. It is believed to be stabilized by reduced dimensionality, non-centrosymmetric crystal structure and large spin orbit coupling~\cite{2020PRLBelitz}. Here, we demonstrate that $5f$-electron system URhSn provides a novel scheme in pressure-temperature phase diagram of correlated electron materials.

URhSn crystallizes in ZrNiAl-type non-centrosymmetric hexagonal structure in which U-atoms lie on a frustrated quasi-kagome lattice in the $c$-plane (Fig. 1a)~\cite{dwight1974UTSn}. Two phase transitions occur in URhSn at temperatures $T_{\rm O}$=52-58 K and $T_{\rm C}$=16-17 K, where  $T_{\rm O}$  is well characterized by a large jump in specific heat (10 J/K-mol), sharp drop in electrical resistivity, and a feeble kink in magnetization~\cite{2020PRBShimizu,Tran_UTSn, mirambet1995physical, palstra1987, Tran_UTSn_1991}. Although the ground state of URhSn is ferromagnetic, the nature of order parameter for the phase transition at $T_{\rm O}$ remains hidden in neutron scattering~\cite{mirambet1995physical} and $^{119}$Sn M\"{o}ssbauer spectroscopy~\cite{kruk1997URhSn_Mossbauer}. A quadrupolar order or a non-collinear magnetic phase below $T_{\rm O}$ is inferred by a reinforcement of $T_{\rm O} (H)$ in magnetic fields parallel to the $c$-axis~\cite{2020PRBShimizu}. A small Sommerfeld coefficient ($\gamma$ = 12 mJ/mol K$^2$) in specific heat capacity and a relatively large ordered moment of 2.1 $\mu_{\rm B}$/U observed in neutron scattering are consistent with a weak $c$-$f$ hybridization and nearly-localized $5f$-electrons in URhSn at ambient pressure~\cite{mirambet1995physical, 2020PRBShimizu}. 

%
%
\begin{figure*}
\includegraphics[width=0.98\textwidth]{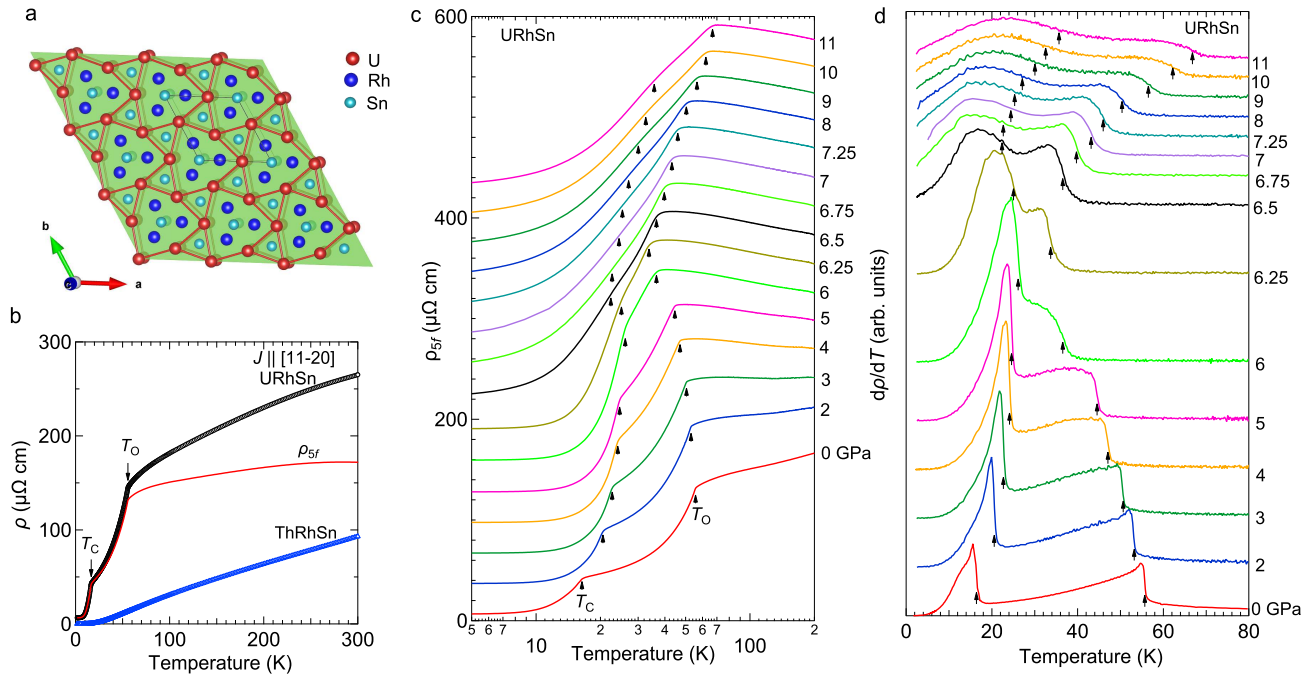}
\caption{\label{Fig1} (a) Crystal structure of URhSn. The triangular units of U atoms constitute a quasi-kagome lattice in the $c$-plane. A sandwiching mirror plane containing Rh and Sn atoms parallel to this quasi-kagome layer is shown with translucent green colour. (b) Electrical resistivity of URhSn for the current along [11$\bar{2}$0]-direction along with of non-magnetic analog ThRhSn. In ThRhSn case the residual resistivity has been subtracted. The red curve is the difference of the two resistivities giving contribution from the $5f$-electrons, i.e. $\rho_{5f}$. (c) Pressure dependence of $\rho_{5f}(T)$ curves in URhSn on logarithmic temperature scale. The $\rho_{5f}(T)$ traces measured under pressure have been shifted vertically by 30 $\mu\Omega$ cm with respect to the adjacent ones for clarity. (d) Temperature derivative of resistivity with pressure in URhSn.  Vertical arrows in (c) and (d) point the transition temperatures (see Fig. S3 in supplementary information for the criteria used in determining $T_{\rm O}$ and $T_{\rm C}$).  }
\end{figure*}
%

Czochralski method equipped in a tetra-arc furnace was utilized to grow the URhSn single crystal. Single crystals were annealed at 800 $^\circ$C under a vacuum better than 10$^{-6}$ torr for 30 days. Although, annealing does not affect the magnetic properties of URhSn, the annealed samples showed sharper transitions at $T_{\rm O}$ and $T_{\rm C}$, as seen by specific heat~\cite{2020PRBShimizu}. The high quality of the single crystal used for resistivity under pressure experiments was confirmed by sharp Laue spots, high RRR (41) and low residual resistivity (6.5~$\mu\Omega$ cm) as well as quantum oscillations of frequencies up to 1.55 kT (see Figs. S1, S2 in supplementary information).  The effective mass for the largest detected Fermi surface in de Haas-van Alphen (dHvA) effect is found to be 1.5 $m_0$, where $m_0$ is rest mass of an electron. We determine the Dingle temperature to be 2.0 K, which furnishes a mean free path of 1020 \AA ~in the URhSn single crystal assuming a spherical shape of the Fermi surface. A large mean free path further attests high quality of the single crystal.  A bar shaped sample of dimension 0.8 mm$\times$0.3 mm$\times$0.2 mm with long direction parallel to [11-20] was obtained by means of Laue method and a spark erosion cutting machine. For estimation of phonon contribution to resistivity, a polycrystalline sample of ThRhSn was prepared by arc-melting process. At ambient pressure, the resistivity of URhSn and ThRhSn were measured using a commercial Quantum Design physical property measurement system. Electrical resistivity under pressure measurements in temperature range 2.5-300 K were performed in a cubic anvil pressure cell~\cite{2004HPR Mori Cubic}. Pressure increments were done monotonically from 2 GPa to 11 GPa at 300 K, at which the pressure transmitting medium Flourinert remains in liquid state. Electrical resistance was measured by a standard four probe technique with a small current excitation of 3 mA passing along the hexagonal [11$\bar{2}$0] crystallographic direction. Electrical contacts on the sample with 20 $\mu$m thin gold wires were made with silver paste.

\begin{figure*}
\includegraphics[width=0.9\textwidth]{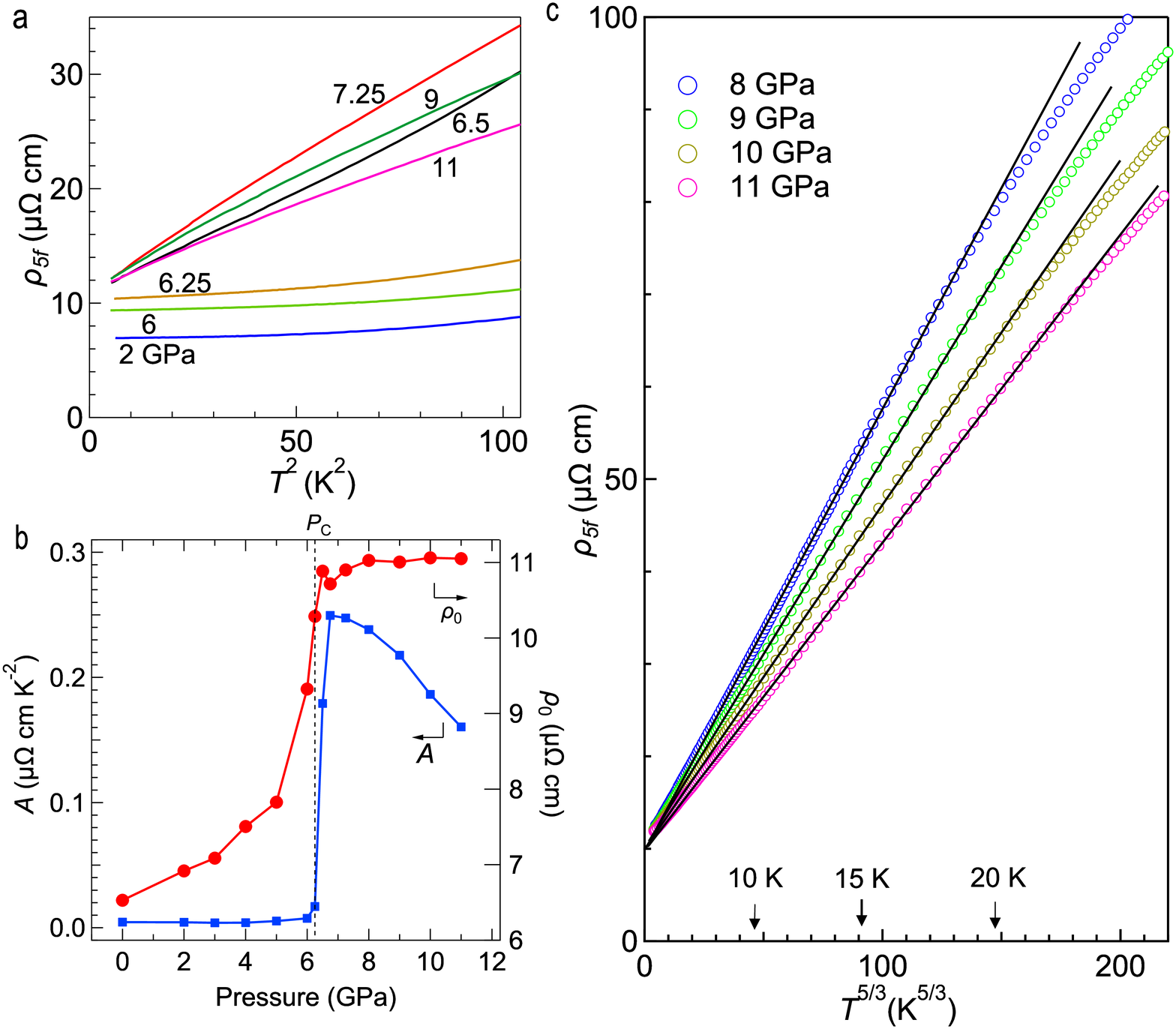}
\caption{\label{Fig2} (a) $5f$-electrons derived resistivity of URhSn on quadratic temperature scale at selected pressures. (b) Variation of the $A$-coefficient (left-axis) and $\rho_0$ derived from the least square linear fit to $\rho_{5f}(T^2)$ curves below 5 K. The error bars in fit results are smaller than size of the symbols. (c) $T^{5/3}$ dependence of $\rho_{5f}$ from 8 GPa to 11 GPa demonstrating a non-Fermi liquid behaviour in URhSn over a large range of temperatures and pressures. Black lines are guide to the eyes to show the agreement. }
\end{figure*}

The double phase transitions in URhSn manifest themselves by kink anomalies followed by rapid drops in resistivity, similar to onset of a magnetic order as shown in Fig. 1b. The contribution of $5f$ electrons in electrical resistivity, $\rho_{5f}$ is obtained by subtracting other contributions mainly composed of scattering off phonons as estimated by the non-magnetic analog ThRhSn. Below 20 K, the change in resistivity of ThRhSn remains smaller than 1 $\mu\Omega$ cm and the subtraction process does not alter the resistivity of URhSn, $\rho$ by more than 2\%.
\begin{figure*}
\includegraphics[width=0.9\textwidth]{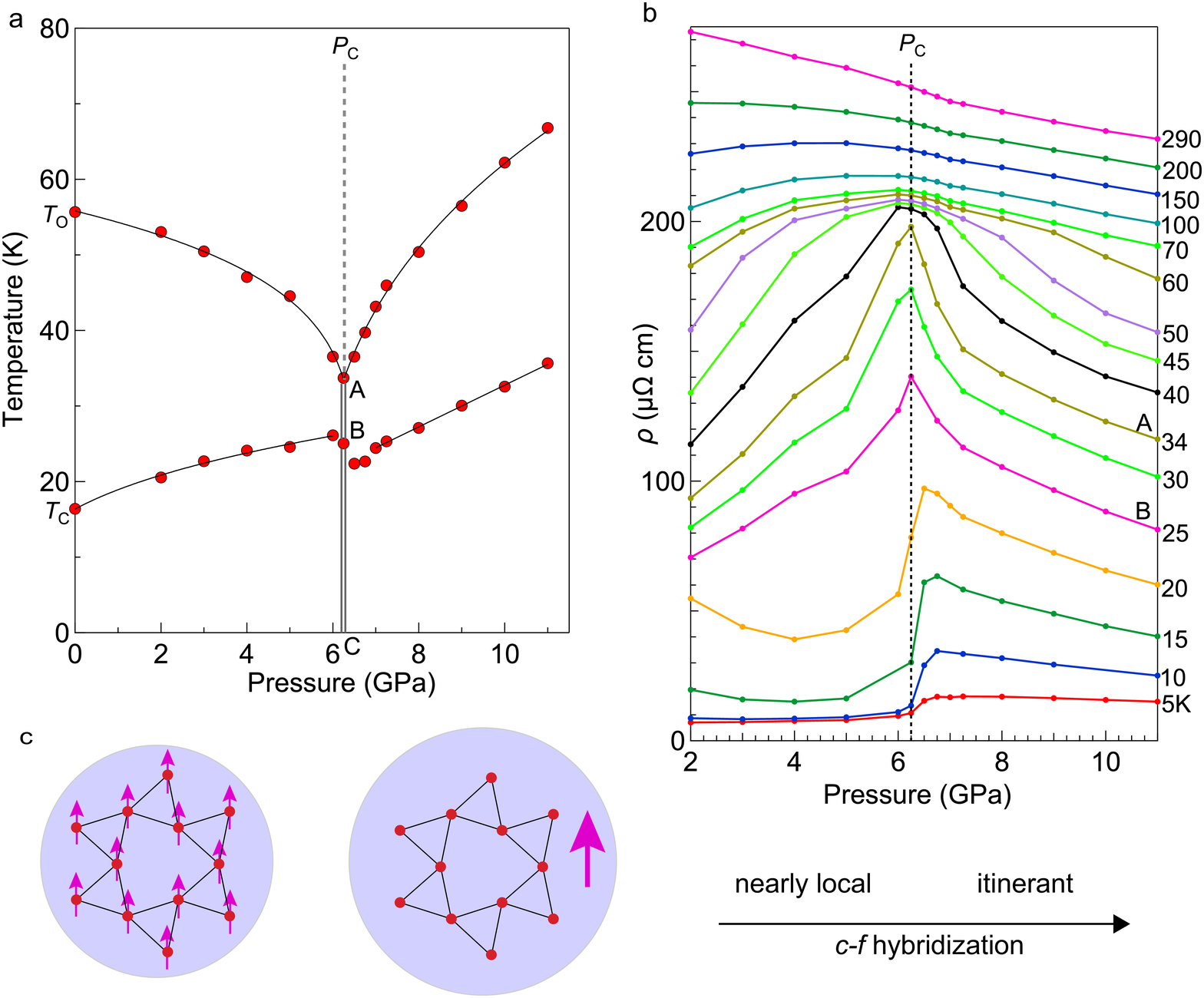}
\caption{\label{Fig2} (a) $P$-$T$ phase diagram of URhSn showing variation of $T_{\rm C}$ and $T_{\rm O}$ with pressure up to 11~GPa. The vertical grey double-line ABC represents a plausible first order phase boundary, which is revealed by sharp anomalies in pressure dependence of resistivity in (b). This first order phase boundary changes into a crossover above A, shown by a dashed line. Putatively, A and B are bicritical points, while C is a quantum phase transition point. (b) Traces showing isothermal variation of resistivity with pressure in URhSn derived from $\rho(T)$ scans. (c) An illustration showing plausible electronic and magnetic ground states in the two pressure regimes across the $P_{\rm C}$. The magnetic electrons attached nearly locally to the quasi-kagome uranium lattice become fully itinerant by pressure induced $c$-$f$ hybridization above the $P_{\rm C}$. The blue enclosures represent size of the Fermi surfaces in the two regimes.}
\end{figure*}

 Fig. 1c shows $\rho_{5f}$ of URhSn measured at various pressures represented on logarithmic temperature scale, assuming that the phonon contribution is unchanged under pressure. The kinks in $\rho_{5f}$($T$) corresponding to the successive phase transitions in URhSn show their presence as steps in temperature derivative (d$\rho$/d$T$), which can be traced up to 11 GPa, albeit above 6 GPa the anomalies at the transition temperatures broaden (Figs. 1c, 1d). On increasing pressure, $T_{\rm O}$ decreases while $T_{\rm C}$ increases, approaching towards each other until a critical pressure, $P_{\rm C}\approx$ 6.25 GPa. The minimum value of $T_{\rm O}(P)$ is 33.7 K at 6.25 GPa, while the maximum value of $T_{\rm C}(P)$ is 26.1 K in the closest vicinity; hence the two transitions remain well separated in the $P$-$T$ plane. A hysteresis or step-like feature in $\rho(T)$ showing a first order behaviour is absent up to 11 GPa. The slope of $\rho_{5f}(T)$  above $T_{\rm O}$ gradually decreases; a -ln$T$ behaviour above 3 GPa is evident, depicting enhancement in the $c$-$f$ hybridization with pressure. 
$\rho_{5f} (T^2)$ below 10 K of URhSn shown in Fig. 2a for selected pressures reveals a remarkable evolution above 6.25 GPa, indicating a drastic change in the electronic structure. The $A$-coefficient derived by fitting equation $\rho_{5f}(T)=\rho_0+AT^2$ between 2.5-5 K shows an abrupt increase from 0.007(5) $\mu\Omega$ cm K$^{-2}$ at 6 GPa to 0.249(6) $\mu\Omega$ cm K$^{-2}$ at 6.75 GPa (Fig. 2b). Concomitantly, the residual resistivity, $\rho_0$ shows a sharp increase and plateau above 6.75 GPa. This together with $\rho_{5f}\propto$ -ln$T$  behaviour strongly indicate a stabilization of a pressure induced phase in which the $5f$-electrons are strongly hybridized with conduction bands. 

 A sub-quadratic behaviour in $\rho_{5f}(T)$ is hallmark of non-Fermi liquid behavior and is ubiquitous in metallic magnets near quantum criticalities~\cite{Hilbert2007RMP, Pfleiderer2009RMP, Stewart 2001 RMP NFL}. $\rho_{5f}$ curves in the pressure induced phase acquire an upward curvature with $T^2$, indicating an emergent non-Fermi liquid phenomenon (Figs. 2a, supplementary information Fig. S5). At and above 8 GPa, a linear variation with $T^{5/3}$ can be readily seen over a substantial temperature range, though a Fermi liquid behaviour is restored at low temperatures (Fig. 2c and supplementary information Fig. S5). A similar exponent has been observed in resistivity of several itinerant electron systems, namely, URhAl~\cite{Shimizu2015URhAl}, U$_3$P$_4$~\cite{2015 Araki U3P4}, Ni$_3$Al~\cite{2005 Niklowitz Ni3Al} and Ni$_x$Pd$_{1-x}$~\cite{1999 Nicklas NiPd} close to the ferromagnetic instability as expected from the self-consistent renormalization (SCR) theory for itinerant electrons in 3D~\cite{Hilbert2007RMP, 1973 Mathon}. The low temperature pressure induced phase above the $P_{\rm C}$ in URhSn is identical to the ferromagnetic state in ZrZn$_2$, in which a $T^{5/3}$ behaviour over an extended region in pressure-temperature ($P$-$T$) plane well below the Curie temperature is attributed to a non-local marginal Fermi liquid state~\cite{2008 Smith ZrZn2, 2012 PRB Sutherland ZrZn2 Marginal, 2007 Takashima  ZrZn2 NFL, 1995 Grosche ZrZn2}. 
The magnetic and electronic phases are represented in the $P$-$T$ phase diagram of URhSn in Fig. 3a. The bicritical points associated with $T_{\rm O}$ and $T_{\rm C}$ are represented by points A (6.25 GPa, 33.7 K) and B (6.25 GPa, 25.1 K), respectively. The line joining A and B extrapolates to a quantum phase transition (QPT) point C at absolute zero temperature. Fig. 3b shows resistivity as a function of pressure, $\rho(P)$ at various temperatures derived from isobaric thermal scans. Onset of step in $\rho(P)$ at $P_{\rm C}$ at low temperatures, which transforms into a sharp peak above B is indicative of a first-order nature of the phase boundary ABC. This pressure-induced first-order phase transformation becomes a crossover above A, which is noticed up to room temperature. Hence, $\rho(P,T)$ of URhSn consistently suggest a QPT at 6.25 GPa between competing magnetic and electronic structures. 

Another interesting aspect in the phase diagram of URhSn is a coupled response of $T_{\rm O}$ and $T_{\rm C}$ with pressure, despite being well separated in temperature and apparently possessing distinctive order parameters. It is noteworthy that a quadratic drop in resistivity between $T_{\rm O}$ and $T_{\rm C}$  gradually linearizes with pressure, with a slope that maximizes at $P_{\rm C}$ (see Fig. S6 in supplementary information). The role of the undetermined order parameter at $T_{\rm O}$ remains an open question in the unusual $P$-$T$ phase diagram of URhSn.

Now we discuss perhaps the most astounding novelty in a $P$-$T$ phase diagram illustrated by URhSn. The nature of $f$-electronic states as well as the associated QPTs have been a matter of long debate. Initial theories of QPTs were stemmed from itinerant electron spin fluctuation models which propose so-called `SDW’ scenario~\cite{1975 Moriya SF, 1976 Hertz Quantum} . It deals with a continuous tuning of long-range spin density wave (SDW) order set in itinerant electrons to absolute zero temperature. This scheme works under framework of SCR theory, which predicts a $T^{5/3}$ behaviour in resistivity by itinerant ferromagnetic fluctuations close to a FM QPT for 3D case. However, lately, this picture was found to be inappropriate for QPTs in some $f$-electron materials, e.g. CeCu$_{5.9}$Au$_{0.1}$ and YbRh$_2$Si$_2$, in which a simultaneous change in Fermi surface takes place~\cite{2008 Gegenwart NatPhys}. To account this new universality class, a `Kondo breakdown’ scenario was proposed~\cite{2010 Si QCP, 2001 Coleman review local QCP, 2001 Si local QCP, 2005 Coleman QCP outlook}, while another approach emphasized valence change~\cite{2007 Miyake}. Note that, in both the cases a pressure induced QPT demarcates a magnetically ordered state from a phase with no magnetic order, in a sharp contrast to the case in URhSn. At ambient pressure, URhSn exhibits nearly local moment magnetism. Under the influence of $c$-$f$ hybridization introduced by pressure, it transforms into a phase that can be suitably described as an itinerant ferromagnet after a jump in the Fermi-surface volume at $P_{\rm C}$ (Fig. 3c). So, the case of URhSn does not fit to either of the abovementioned classifications and demands a new theoretical formulation. An interplay of competing orders with geometric or magnetic frustration inherent in the quasi-kagome lattice is a potential cause for the unique magnetic and electronic instability in this $5f$-electron system. Our results show that quantum phase transition in $f$-electron systems is a more general phenomena than previously thought and may involve multicritical  points.

 We thank V. Taufour, M. Yokoyama, Y. Tokunaga, H. Harima and Y. \={O}nuki for fruitful discussions. 
  We thank S. Nagasaki for her technical support during high pressure experiments. 
  We acknowledge financial support from JSPS KAKENHI grant numbers JP18F18017, JP19H00648, JP15H05882, JP15H05886, JP15K21732, JP15H05884, JP19H00646, JP19J20539, JP17K14328, and JP20K03851.

\end {document}